\documentclass{elsart1p}
\usepackage{dcolumn}
\usepackage{bm}
\usepackage{mathrsfs}
\usepackage{amsmath}
\usepackage{epsfig}
\journal{Physica A}
\begin{document}
\begin{frontmatter}
\title{Glassiness in Simple Liquids}
\author[]{Ziv Rader\corauthref{*}},
\corauth{Corresponding author.}
\ead{zivrader@post.tau.ac.il}
\author[]{Moshe Schwartz}
\address{Raymond and Beverly Sackler Faculty of Exact Sciences, School of Physics and Astronomy,\\Tel Aviv University, Ramat Aviv, Tel Aviv 69978, Israel}
\begin{abstract}In previous work the parameter of glassiness was introduced to distinguish between a liquid and a glass, using a formal analogy with the quantum Bose system. The glassiness is defined in such a way that it is unity in a frozen system and less than one in a liquid. In the present letter we revise first the results obtained for the glassiness in a hard sphere liquid as a function of the density. Then we investigate the influence of an attractive potential by obtaining the glassiness as a function of the density, temperature and the attractive tail when a square well  potential is added to the hard core.
\end{abstract}
\end{frontmatter}
\newcommand{\D}{\mathrm{d}}
\newcommand{\R}{\mathbf{r}}
\newcommand{\kb}{k_B}
The transition from the liquid state to the glassy state can not be seen as a change in an equilibrium parameter, as  when dealing with the liquid-solid phase transition. Transport parameters such as the diffusion constant of an external probe and the viscosity are the main criteria for distinguishing the glassy state from a simple classical liquid \cite{Bib12,Bib13}. Some years ago an analogy between a simple liquid and a Bose quantum system was used to suggest an equilibrium parameter, which distinguishes between a simple liquid and a glass [1]. That parameter is based on the difference in the symmetry under permutation of the classical liquid and the glass. In both states the system is invariant under permutations but still there is an important difference in the physical effectiveness of the symmetry. In the liquid state configurations which differ by permutations are connected by trajectories in configuration space, which do not have to cross high energy barriers. In the glass, on the other hand, such trajectories have to cross extremely high barriers. In quantum systems the Bose condensed fraction is the parameter that measures the effectiveness of the symmetry under permutations. In the following we present the analogy between a classical system at equilibrium and a quantum Bose system and show that a condensed fraction ,$\xi$, can be defined, which is non zero in the liquid state and zero in the glass, where the symmetry under permutations is not relevant. The glassiness , $\Gamma$, is defined as
\begin{equation}
\label{A0}
\Gamma=1-\xi.
\end{equation}

\begin{figure}[!h]
\centering 
\graphicspath{{Figures/}}
\epsfig{file=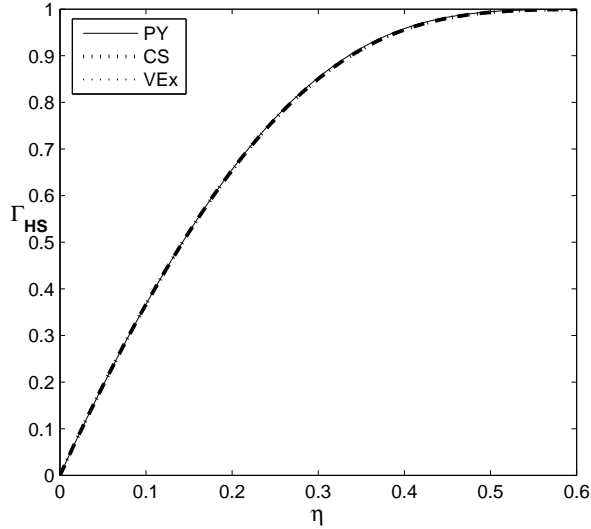,width=0.65\textwidth}
\caption{Glassiness of the HS system in 3D.} 
\label{Fig1}
\end{figure}

\begin{figure}[!h]
\centering 
\graphicspath{{Figures/}}
\epsfig{file=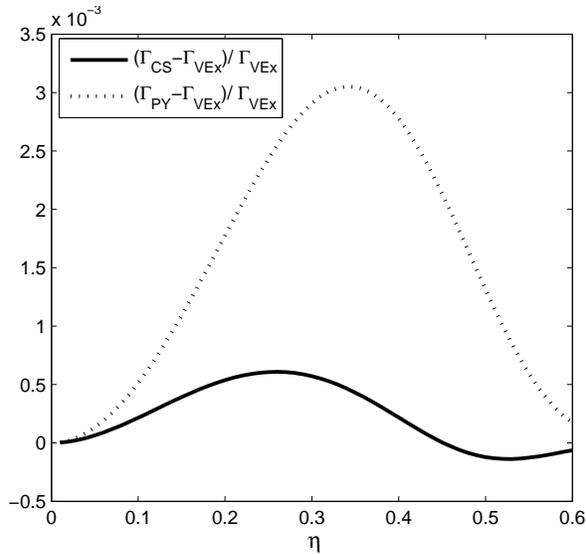,width=0.65\textwidth}
\caption{Relative differences of the glassiness calculated from the three different EOS.} 
\label{Fig4}
\end{figure}

We begin with a set of $N$ coupled Langevin equations for a classical system of volume $V$.
\begin{equation}\label{A1}
\gamma\dot{\mathbf{x}}_i=-\frac{\partial
W}{\partial\mathbf{x}_{i}}+\eta_{i},
\end{equation}
Where $W$ is the interparticle potential and $\eta_i$ is a noise experienced by the $i$'th particle, obeying 
\begin{equation}\label{A2}\begin{array}{lllll}
\langle\eta_{i}\rangle=0&& {\rm and} &&
\left\langle\eta_{i}\left(t\right)\eta_{j}\left(t'\right)\right\rangle=D\delta_{ij}\delta(t-t').
\end{array}
\end{equation}
The above provides a framework which allows a natural analogy between the equilibrium state of the classical system and the ground state of a corresponding quantum system. The Fokker-Planck equation for the distribution ,$\mathcal{P}$, that corresponds  to the above set of Langevin equations  has an equilibrium solution $\mathcal{P}_s=exp[-W/\kb T]$, where $\kb T=D/2\gamma$ \cite{Bib8}. The standard transformation $\mathcal{P}=\mathcal{P}_s^{1/2}\Psi$, yields  an imaginary time Schr\"{o}dinger equation
for $\Psi$,
\begin{equation}\label{A3}
\frac{\partial \Psi}{\partial t}=-\mathcal{H}\Psi.
\end{equation}
The Hamiltonian above is a non-negative definite Hermitian operator and its only state with eigenvalue zero is the ground state, $\Psi_G=\mathcal{P}_s^{1/2}$. Because it is symmetric under permutations it is also the ground state of the Bosonic reduction of
$\mathcal{H}$. The state $\mathcal{P}_s^{1/2}$ is thus a natural Bose "quantum" ground state that corresponds to the classical system and the question of the value of the condensed fraction becomes a legitimate question for the classical system, which as discussed above has bearing on the question whether the system is a liquid or a glass. The purpose of the present article is to study  the glassiness in the liquid state for the hard sphere system and then consider the effect of adding an attractive interaction. For a potential energy which is the sum of two body interactions $\phi(r_i-r_j)$, the ground state wave function is given by
\begin{equation}\label{A5}
\Psi_G=exp\bigg[-\frac{\beta}{4}\sum_{i\neq j=1}^N\phi(\mathbf{x}_i-\mathbf{x}_j)\bigg].
\end{equation}
where $\beta=1/\kb T$. Some basic algebra was used in ref. [1] to obtain the condensed fraction,  
\begin{equation}\label{A6}
\xi=\frac{Q_{N+1}}{VQ_N}\frac{G(\beta/2)}{G(\beta)},
\end{equation}
where $Q_N$  is the configurational  partition function of a system of $N$ particles and the function $G(\alpha)$  is defined by 
\begin{equation}\label{A7}
G(\alpha)=\left\langle exp\left[-2\alpha\int\D\mathbf{x}'\phi(\mathbf{x}')\rho(\mathbf{x}')\right]\right\rangle,
\end{equation}
where $\rho(\mathbf{x})=\sum_{i=1}^N\delta(\mathbf{x}-\mathbf{x_i})$ and $\langle\cdots\rangle$ denotes thermal average with respect to $\mathcal{P}_s$. The volume limit is implied in eq.(\ref{A7}) .
For the hard sphere case  the $G$ factors cancel, so that the condensed fraction can be easily expressed in terms of the chemical potential. We begin with the following
relation
\begin{equation}\label{A8}
\frac{Q_N}{V^N}=e^{-\beta F_{ex}},
\end{equation}
where $F_{ex}$ is the configurational part of the free energy and is given by
\begin{equation}\label{A9}
F_{ex}=\frac{N}{\beta}\int_0^\eta\bigg{(}\frac{P(\eta^{'})V}{N\kb T}-1\bigg{)}
\frac{\D\eta^{'}}{\eta^{'}},
\end{equation}
where $\eta$ is the packing fraction which is given  in terms of the density $\rho$  and the hard sphere diameter $\sigma$  as $\eta=\pi\sigma^3\rho/6$ and $P$ is the pressure. The excess chemical potential associated with this free energy is
\begin{equation}\label{A9A}
\mu_{ex}=\frac{1}{\beta}\int_0^\eta\bigg{(}\frac{P(\eta^{'})V}{N\kb T}-1\bigg{)}
\frac{\D\eta^{'}}{\eta^{'}}.
\end{equation}
The glassiness of the hard sphere system, just like all other thermodynamic properties is independent of the temperature and is given by 
\begin{equation}\label{A9B}
\Gamma_{HS}=1-\left(\frac{Q_{N+1}}{V^{N+1}}\right)\left(\frac{V^N}{Q_N}\right)=1-exp\left\{\mathop{\lim}\limits_{N\to\infty}\left[-\beta\left(F_{ex}^{N+1}-F_{ex}^N\right)\right]\right\},
\end{equation}
so that
\begin{equation}\label{A10}
\Gamma_{HS}=1-exp(-\beta\mu_{ex})=1-exp\left[-\int_0^\eta\left(\frac{P(\eta^{'})V}{N\kb T}-1\right)
\frac{\D\eta^{'}}{\eta^{'}}\right].
\end{equation}
The equation of state for the hard sphere system has been obtained in the literature by numerous
analytical approximations such as Percus-Yevick (PY) \cite{Bib2}, the Carnahan-Starling (CS) \cite{Bib3}, the virial expansion (VEx)\cite{Bib4} etc. All of the above yield analytic expressions for the dependence of the pressure on the packing fraction which lead directly to analytic expressions for the glassiness. For the virial expansion the expression is trivial, of course. The Percus-Yevick and Carnahan-Starling produce respectively the following equations of state: 
\begin{subequations}
	\begin{equation}
	\left(\frac{P V}{N\kb T}\right)_{PY}=\frac{1+\eta+\eta^2}{(1-\eta)^3},
	\end{equation}
	\begin{equation}
	\left(\frac{P V}{N\kb T}\right)_{CS}=\frac{1+\eta+\eta^2-\eta^3}{(1-\eta)^3},
	\end{equation}
\end{subequations}
which lead to the following expressions for the glassiness respectively,
\begin{equation}\label{A11}
\Gamma_{_{PY}}=1-\left(1-\eta\right)e^{\frac{3}{2}\left[1-1/\left(1-\eta\right)^2\right]}
\end{equation}
and
\begin{equation}\label{A12}
\Gamma_{_{CS}}=1-e^{-\frac{(4-3\eta)\eta}{(1-\eta)^2}}.
\end{equation}
Figure \ref{Fig1} presents the glassiness of the HS system, following from the approximations considered above. As can be seen from figure \ref{Fig1}, it is not easy to distinguish among the various approximations for the glassiness. In figure \ref{Fig4} we take the VEx glassiness as a reference and present the relative departure of the PY and CS glassiness from the reference.
\\

We note that the results obtained above for the classical hard sphere system can also be used to estimate the condensed fraction in liquid $He^4$ under various condition. This observation is based on the work of Penrose and Onsager \cite{Bib7} who used the classical hard sphere weight function as a model of the ground state function of liquid $He^4$. They used that wave function to obtain the condensed fraction at 0.28 of the close packing density by employing the low density expansion to its lowest order.
Our present calculation goes beyond the lowest order and supplies therefore, the condensed fraction in $He^4$ in a wider range of densities accessible under pressure, within the Penrose Onsager model.
\\

\begin{figure}[!h]
\centering
\graphicspath{{Figures/}}
\epsfig{file=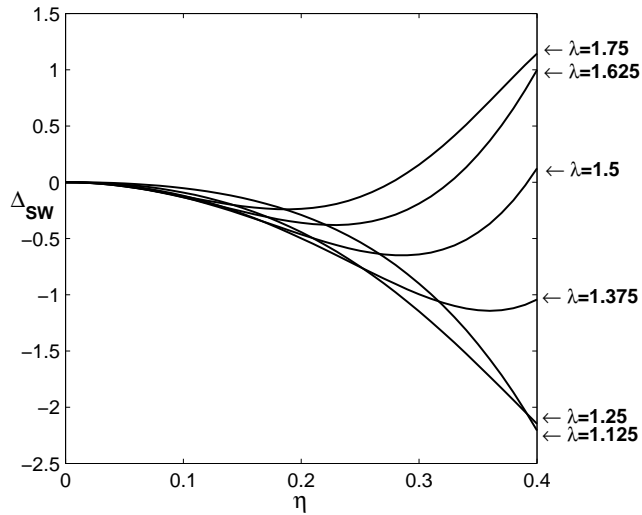,width=0.65\textwidth}
\caption{$\Delta_{SW}$ as a function of the packing fraction, $T^*=1$.}
\label{Fig2}
\end{figure}

\begin{figure}[!h]
\centering
\graphicspath{{Figures/}}
\epsfig{file=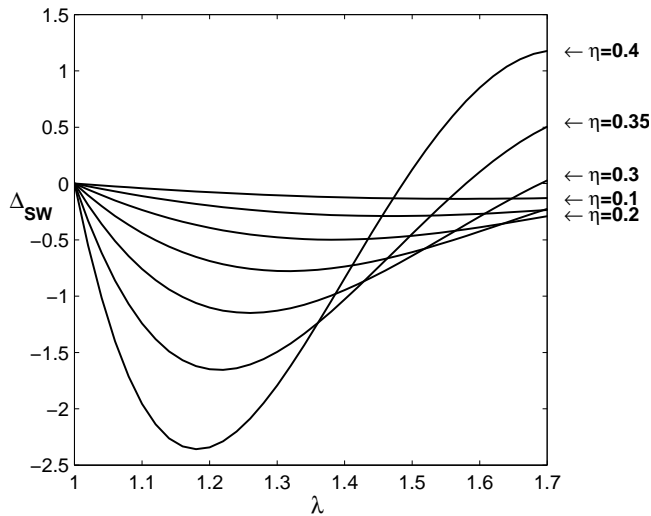,width=0.65\textwidth}
\caption{$\Delta_{SW}$ as a function of the well width, $T^*=1$.}
\label{Fig3}
\end{figure}

The hard sphere system is a canonical model used in the study of simple liquids. Yet, realistic potentials have an attractive tail, which must affect the physical properties of the system including the glassiness. The effect of an attractive tail seems unclear. On one hand we expect it to counteract the repulsive potential and thus the introduction of an attractive tail seems to lead to a reduction of the glassiness. The addition of an attractive potential is expected on the other hand to increase the order in the system and is expected thus also to increase the glassiness. The overall effect seems to be density dependent. At low densities the first effect is expected to dominate, because introduction of an attractive tail will not increase the tendency to order and the glassiness will decrease. At high densities the latter effect will prevail. To get a basic idea of what is going on we add to the hard core a soft attractive potential. The two body potential is thus
\begin{equation}\label{C1}
\phi=\phi_{HS}+\phi_{SP}.
\end{equation}
To first order in the soft potential the glassiness is given by
\begin{equation}\label{C3}
\Gamma=\Gamma_{HS}+(1-\Gamma_{HS})\Delta_{SP},
\end{equation}
where the dimensionless quantity, $\Delta_{SP}$, is the first correction to the condensed fraction divided by the condensed fraction of the hard sphere system, which is just the first order correction to $\ln{\xi}$. We can calculate the first order correction to $\ln{\xi}$ by considering separately $\ln{(Q_{n+1}/V Q_N)}=-\beta\mu_{ex}$ and $\ln{G(\beta/2)/G(\beta)}$. The first order correction to the first term was obtained in the past \cite{Bib15,Bib16} and is given by
\begin{equation}\label{C3B}
\Delta_{SP}^1=-\frac{\beta}{2}\frac{\partial}{\partial\rho}\left[\int\rho^2 g_{_{HS}}(r,\rho)\phi_{SP}(r)\D r\right],
\end{equation}
where $g_{_{HS}}(r,\rho)$ is the radial distribution function of the hard sphere system at density $\rho$, which is assumed to be given. We obtain the first order correction to the second term to be given by 
\begin{equation}\label{C3C}
\Delta_{SP}^2=\beta\rho^2\int g_{_{HS}}(r,\rho)\phi_{SP}(r)\D r.
\end{equation}
Thus, to first order in the soft potential the total correction is given by 
\begin{equation}\label{C4}
\Delta_{SP}=\frac{\beta\rho^2}{2}\int\frac{\partial
g_{_{HS}}(r,\rho)}{\partial\rho}\phi_{SP}(r)\D r.
\end{equation}
A common idea used in the theory of classical liquids which is considered to give a fairly accurate description of the liquid is to use the simplest attractive potential which is the square well potential defined by
\begin{equation}\label{C2}
\phi_{SW}(\mathbf{r})=\left \{ \begin{array}{lll} 0,&&
|\mathbf{r}|<\sigma\\-\epsilon, && \sigma<|\mathbf{r}|<\lambda
\sigma\\0,&&\lambda \sigma<|\mathbf{r}|.
\end{array}\right .
\end{equation}
For the SW potential the expression (\ref{C4}) can be simplified:
\begin{equation}\label{C5}
\Delta_{SW}=\frac{-\epsilon\beta\rho^2}{2}\int_{\sigma}^{\lambda\sigma}\frac{\partial
g_{_{HS}}(\mathbf{r},\rho)}{\partial\rho}\D \mathbf{r}.
\end{equation}
In order to evaluate this correction, we use an expansion in the density of the pair distribution function,
\begin{equation}\label{C6}
g_{_{HS}}(r,\rho)=1+\sum_{n=1}^{\infty} H_n(r/\sigma)(\rho\sigma^3)^n\hspace{0.4cm}for\hspace{0.2cm}r>\sigma.
\end{equation}
This density expansion was obtain on the basis of some analytical work and MC integrations up to six's order \cite{Bib5} and is hereby used to calculate the integral in (\ref{C5}) (There are other forms that give the radial distribution function and the structure factor as a function of the density \cite{Bib6,Bib9,Bib10,Bib11} but the form given by (\ref{C6}) proved to be the most convenient for our purpose). This results in explicit dimensionless coefficients $a_n(\lambda)$ for $n=1-6$ in the expansion of $\Delta_{SW}$ in the packing fraction $\eta$,    
\begin{equation}\label{C7}
\Delta_{SW}=-\frac{12}{T^{\star}}\sum_{n=1}^{6} a_n(\lambda)\eta^{n+1},
\end{equation} 
where $T^{\star}=1/\beta\epsilon$ is a reduced temperature. The correction, $\Delta_{SW}$, for $T^{\star}=1$ is presented in figure \ref{Fig2} as a function of the packing fraction for different well widths and in figure \ref{Fig3} as a function of the well width for different packing fraction values.
We see that the results presented in the figures agree with our previous expectations. The correction decreases first as the density is increased reflecting the fact that the attractive potential counteracts the hard core. Then the correction starts to increase as the density is further increased, because it supports the generation of local order. We see that the packing fraction at which the correction starts to increase strongly depends on the value of $\lambda$ (for the smallest $\lambda$'s it does not start to increase). Indeed, such a behavior also agrees with our previous consideration, because the larger $\lambda$ the more effective is the attractive potential in supporting local order, so, if the increase of the correction is due to the generation of local order, it must increase first for larger value of $\lambda$.
\\

In summary, we have calculated the glassiness in simple liquids described either by a hard sphere interaction or by a hard sphere to which a weak square well potential has been added. For the hard sphere system we find in the whole range of the packing fraction between 0 and close packing that the glassiness curves, obtained from three different equations of state for the hard sphere system, are practically identical. Clearly, we do not expect our results to be valid in all that range because of the simple fact that the PY and CS equations of state are only approximate and can not expected to hold beyond $\eta=0.5$. The effect of a weak attractive potential depends trivially on the dimensionless strength of the potential (the reduced temperature). Its dependence on the packing fraction $\eta$ and the ratio $\lambda$ between the square well and the hard sphere radii is more interesting. Essentially, if both $\eta$ and $\lambda$ are small the glassiness is reduced. The reason is that the attractive potential just counteracts the hard sphere. When both parameters are large the glassiness is enhanced because, the attractive potential supports generation of  short range order in the system. When one of the parameters is small and the other is large, the resulting glassiness depends on detail but our general argument which is supported by the figures we present, suggests that fixing one of the parameters and increasing the other (the increase of $\eta$ is limited, of course) will eventually lead to an increase of the glassiness.

\end{document}